\documentclass[onecolumn,manuscript,amsmath,amssymb,pra]{revtex4-1}
\usepackage{graphicx}
\usepackage{graphics}
\usepackage{color}
\usepackage{rotating}
\usepackage{array}
\usepackage{slashbox}
\usepackage{bm}
\usepackage{float}

\usepackage[english]{babel}
\usepackage[T1]{fontenc}
\usepackage[latin1]{inputenc}
\usepackage{subfigure}
\usepackage{footnote}
\usepackage{multirow}
\usepackage{amsmath}
\usepackage{amssymb}
\usepackage{tabularx}
\usepackage[margin=1in]{geometry}       
\usepackage{amsthm}
\usepackage{tikz}

\def \i{\`{\i}~}

\def     \i{\`{i}~}

\begin{document}

\title{ ~~~~~~ Particle capture by drops in turbulent flow}

\author{
~\\
A. Hajisharifi}

\affiliation{
Department of Engineering and Architecture, University of Udine, 33100 Udine, Italy}
\author{
C. Marchioli}
\email{Corresponding Author: marchioli@uniud.it}
\affiliation{
Department of Engineering and Architecture, University of Udine, 33100 Udine, Italy}
\affiliation{
Department of Fluid Mechanics, CISM, 33100 Udine, Italy
}
\author{
A. Soldati}
\affiliation{
Institute of Fluid Mechanics and Heat Transfer, TU Wien, 1040 Wien, Austria}
\affiliation{
Department of Engineering and Architecture, University of Udine, 33100 Udine, Italy}

\date{\today}
\vspace{0.5cm}
%


\begin{abstract}
{\bf ABSTRACT}\\
We examine the process of particle capture by large deformable drops in turbulent channel flow.
We simulate the solid-liquid-liquid three-phase flow with an Eulerian-Lagrangian method based on Direct
Numerical Simulation of turbulence coupled with a Phase Field Model, to capture the interface dynamics,
and Lagrangian tracking of small (sub-Kolmogorov) particles.
Drops have same density and viscosity of the carrier liquid, and neutrally-buoyant solid particles
are one-way coupled with the other phases. Our results show that particles are transported towards
the interface by jet-like turbulent motions and, once close enough, are captured by interfacial forces
in regions of positive surface velocity divergence.
These regions appear to be well correlated with
high-enstrophy flow topologies that contribute to enstrophy production via vortex compression or stretching.
Examining the turbulent mechanisms that bring particles to the interface, we have been able to derive a
simple mechanistic model for particle capture. The model can predict the overall capture efficiency and
is based on a single turbulent transport equation in which the only parameter scales with the turbulent kinetic
energy of the fluid measured in the vicinity of the drop interface. The model is valid in the limit of non-interacting particles and
its predictions agree remarkably well with numerical results.
\end{abstract}
\maketitle

\section{Introduction}

The process of particle capture by drops or bubbles in a turbulent flow is of relevance in a number of industrial applications requiring particulate abatement, e.g. via wet scrubbing \cite{kim2001particle, peukert2001industrial, meikap2002scrubbing}. The very same process is also observed in environmental problems, such as accidental oil spills in which oil interacts with sediments to form oil-particle aggregates that may affect the transport of spilled oil and enhance oil biodegradation \cite{zhao2017isolation}.
In these applications, particle capture occurs in two steps: First, particles move towards the drop/bubble surface under the influence of turbulence in the carrier liquid, possibly aided by external forces as in the case of electrostatic scrubbing \cite{di2015capture,su2020experimental}; then particles stick the drop/bubble surface upon inertial impaction or turbulent diffusion. Particle behaviour upon impaction determines the overall attachment efficiency, which in turn affects the overall capture efficiency of the drop/bubble. In this context, a crucial physical property is surface tension, which controls drop/bubble deformability, drives particle adhesion and leads to the formation of a layer that may change the mechanical and mass transport properties of the interface \cite{dinsmore2002colloidosomes,dickinson2010food,stratford2005colloidal}.

Aim of the present study is to elucidate the physical mechanisms that govern particle capture at the surface of a swarm of deformable drops transported by a turbulent flow, focusing in particular on the characterization of the flow events that bring particles to the surface. This interest is motivated by the need of detailed information about the near-interface flow field in usual engineering practice, e.g. for the development of physics-based models or correlations able to predict transfer rates across a liquid-liquid interface \cite{goniva2010}. Currently, industrial CFD tools can only rely on mechanistic correlations to predict the capture efficiency of a
full-scale equipment \cite{goniva2010,rafidi2018cfd}.

To study the targeted three-phase turbulent flow, we use a computational approach that couples for the first time the direct numerical simulation (DNS) of the carrier fluid and drops with an interface-capturing method for the evolution of the drop surface and a Lagrangian tracking method for the particle dynamics.

Three-phase computational models like the one adopted here pose computational challenges in terms of modelling the interactions among the different phases and the complex dynamics produced by a moving, deformable interface \cite{elghoARFM}. drops introduce additional physical mechanisms into the flow due to their ability to deform but also to breakup and coalesce with other drops, thus changing the overall surface area available for particle capture. The problem is complicated further by the wide range of length scales involved, from the interface thickness - $\mathcal{O}(10^{-9})~m$ - to the particle size - $\mathcal{O}(10^{-5})~m$ - to the drop size - $\mathcal{O}(10^{-2})~m$. Because of these complexities, most of the numerical studies available in the literature focus on the role that surface physicochemical forces have in determining particle adsorption in no-flow or viscous flow conditions, when particle-drop interactions are not affected by the flow hydrodynamics. Examples include the study of the behaviour of a single particle trapped at a planar fluid interface \cite{binks2002particles,park2014particles}, the surface stress tensor modification for a pendant drop covered by a monolayer of particles in the low-Reynolds-number limit \cite{gu2016direct}, or the attachment of a colloidal particle to the surface of an immersed bubble rising in still fluid \cite{lecrivain2016direct}, to name a few recent works. Also relevant is the study by 
\cite{pozzetti2018multiscale}, who developed a DEM-VOF method to reproduce drop formation and interface perturbations from a single particle. The same methodology was applied by
\cite{sun2015three} to study gas-solid-liquid-flows of relevance for sedimentation problems.


All of the above-mentioned studies have contributed to the physical understanding of particle-laden fluid interfaces, but do not consider turbulent flow conditions: Clearly, the flow hydrodynamics must be accounted for in turbulent systems, which are the focus of our investigation and (to the best of our knowledge) have not been examined before. Intuitively, one expects particles to be brought in the near-interface region by coherent jet-like fluid motions able to generate local deformations of the drop surface along the surface itself (via the tangential stress they generate) but also along the interface-normal direction (via the pressure fluctuations and normal stresses they induce). When strong enough, these deformations will produce a change in the topology of the flow surrounding each drop, as compared to the topology of an unladen flow \cite{dodd2019small}, and will play a role in the particle adhesion process. To examine this role, we have concentrated our analysis on the fluid motions that occur in the proximity of the interface, where the smallest hydrodynamic length scales are typically located. For this purpose, we consider a density-/viscosity-matched flow that
allows uncoupling of inertial effects associated with particle size from those due to differential density \cite{abbas}: Even in this simplified case, the presence of an interface is crucial as it represents an elastic, compliant boundary that can modulate the overall energy and momentum transfer the carrier phase and the drops \cite{soligo2020effect}.

The paper is structured as follows. In Sec. \ref{methodology}, the physical problem and the numerical methodology are presented: Specifically, we investigate the interaction between sub-Kolmogorov particles and super-Kolmogorov drops in a channel flow configuration, considering particles with low, albeit different, inertia. In Sec. \ref{results}, first we characterize the flow topology near the locations of the interface at which particles get captured, by means of classical topology indicators \cite{chong1990general}.

 Then, we examine the time evolution of the fraction of captured particles, proposing a simple predictive model to estimate the capture rate.
Finally, in Sec. \ref{conclusions} the main findings are summarized and future perspectives are provided.

\section{Physical problem and methodology}
\label{methodology}

The physical problem considered in this study consists of turbulent three-phase channel flow in which large, deformable drops and small, spherical particles are transported by a carrier liquid. To numerically simulate this flow, we performed Direct Numerical Simulation (DNS) of the Navier-Stokes equations to provide an accurate representation of turbulence, using a Phase-Field Method (PFM) to describe the dynamics of the drop surface (referred to as interface hereinafter) via the Cahn-Hilliard equations and a Lagrangian approach based on a suitably-simplified version of the Maxey-Riley-Gatignol equation to compute the trajectory of the particles. Note that the Navier-Stokes equations, which define the hydrodynamics of the system, include an additional force term to account for the presence of the interface.
In the following, the PFM is presented first and then the force coupling with the Navier-Stokes equations is described. Finally, the Lagrangian particle tracking is discussed.

\subsection{Modeling of the interface dynamics}
The phase field method adopted in this study uses a scalar field (order parameter) to describe the transport of the phase field $\phi$, which provides the instantaneous shape and position of the interface. The phase field is constant in the bulk of both the carrier phase ($\phi= - 1$) and the drops ($\phi= - 1$), while undergoing a smooth transition across the interfacial layer. The position of the interface is given by the iso-level $\phi= 0$. The time evolution of the order parameter $\phi$ is given by the Cahn-Hilliard equation, which reads in dimensionless form as 

\begin{equation}
\frac{\partial \phi}{\partial t}+{\bf{u}}\cdot\nabla\phi=\frac{1}{Pe_{\phi}}\nabla^2 \mu_{\phi}  + f_p~~~,
\label{spfm}
\end{equation}

where ${\bf{u}}=(u,v,w)$ is the carrier fluid velocity field vector, $Pe_{\phi}$ is the  P$\acute{\mathrm{e}}$clet number (ratio between the diffusive and convective time scale controlling the interface relaxation), $\mu_{\phi}$ is the chemical potential and $f_p$ is a penalty flux (not accounted for in the standard phase field formulation) introduced to force the interface toward its equilibrium by reducing the diffusive fluxes induced by the gradient of the chemical potential $\mu_{\phi}$ \cite{li2016phase, zhang2017flux, soligo2019mass}:
Through this term, the equilibrium interfacial profile can be maintained and the drawbacks of the standard formulation (e.g. mass leakage) can be overcome. The penalty flux is defined as:

\begin{equation}
f_p=\frac{\lambda}{Pe_{\phi}} \biggl[\nabla^2\phi - \frac{1}{\sqrt{2} Ch} \nabla \cdot \biggl((1-\phi^2) \frac{\nabla \phi}{|\nabla\phi|} \biggr)  \biggr]~~~,
\label{penaltyFlux}
\end{equation}

with the parameter $\lambda$ set according to the scaling proposed by \cite{zhang2017flux, soligo2019mass}.
The chemical potential $\mu_{\phi}$ is defined as the variational derivative of the Ginzburg-Landau free energy functional, $\mathcal{F}[\phi, \nabla \phi]$. The functional composed as the sum of two different contributions:

\begin{equation}
\mathcal{F}[\phi, \nabla \phi] = \int_{\Omega} ( f_0 + f_m) ~ \mathrm{d} \Omega~~~,
\label{FreeEnergy}
\end{equation}

where $\Omega$ is the reference domain. The two contributions are defined as follows \cite{scarbolo2015coalescence,roccon2017viscosity}

\begin{equation}
f_0=\frac{1}{4}(\phi-1)^2 (\phi+1)^2~~~.
\label{DWP}
\end{equation}

\begin{equation}
f_m= \frac{{Ch}^2}{2} {|\nabla \phi|}^2~~~.
\label{ME}
\end{equation} 

The terms $f_0$ and $f_m$ are the functions of phase field $\phi$. The former, $f_0$, is the double well potential that describes the tendency of the system to separate into two pure fluids and is defined and the latter, $f_m$,  is the mixing energy and accounts for the energy stored at the interface (i.e. the surface tension); $Ch$ is the Cahn number, representing the dimensionless thickness of the interfacial layer. The chemical potential is thus obtained as 

\begin{equation}
\mu_{\phi}= \frac{ \mathcal{F}[\phi, \nabla \phi]}{\delta \phi}=\phi^3 - \phi - Ch ^2 \nabla ^2 \phi~~~.
\label{ChemicalPotential}
\end{equation} 

When the system is at equilibrium, the chemical potential is uniform over the entire domain. The equilibrium profile for a flat interface located at $s=0$, $s$ being the coordinate normal to the interface, can be obtained by solving $\nabla \mu_{\phi} =0$, which yields the following hyperbolic tangent profile

\begin{equation}
\phi_{eq} (s)= \tanh \biggl(  \frac{s}{\sqrt{2} Ch}  \biggr)~~~,
\label{EqProf}
\end{equation} 

which ensures a smooth transition between the limiting values $\phi=\pm 1$ that are reached in the bulk of each phase.
 
\subsection{Hydrodynamics}
The hydrodynamics of the three-phase flow is described by the Continuity and Navier-Stokes equations, which is coupled with the Cahn-Hilliard equation previously introduced. This computational model can handle non-matched properties in general cases \cite{ding2007diffuse,roccon2017viscosity}; density and viscosity can be defined as a function of the phase field $\phi$, but
In this study we assume that the two Eulerian phase, namely the carrier fluid (denoted by subscript $f$) and the drops (denoted by subscript $d$), have matched density $(\rho=\rho_f=\rho_d)$ and matched viscosity $(\eta=\eta_f=\eta_d)$. According to this one-fluid formulation, the dimensionless Continuity and Navier-Stokes equations read as

\begin{equation}
\nabla \cdot {\bf u} = 0~~~,
\label{Mass}
\end{equation} 

\begin{equation}
\frac{\partial {\bf u}}{\partial t}+{\bf u} \cdot \nabla {\bf u}=-\nabla p+\frac{1}{Re_\tau}\nabla^2 {\bf u}+\frac{Ch}{We}\frac{3}{\sqrt{8}}\nabla \cdot\tau_c ~~~,
\label{NS}
\end{equation}

where $\nabla p$ is the pressure gradient, which includes both the mean pressure gradient that drives the flow and the fluctuating part; $Re_{\tau}= u_{\tau} h / \nu_f$ is the friction Reynolds number (based on the friction velocity $u_{\tau}=\sqrt{\tau_w/\rho_f}$, with $\tau_w$ the mean wall shear stress, the channel half height $h$ and the fluid kinematic viscosity $\nu_f$);$We=\rho_f u_{\tau}^2 h / \sigma$ is the Weber number, based on the surface tension $\sigma$ of a clean interface;
 $\tau_c=|\nabla \phi|^2 I - \nabla \phi \otimes \nabla \phi $  is the Korteweg stress tensor, which accounts for the interfacial force induced on the flow by the occurrence of capillary phenomena due to non-local molecular interactions at the interface on the two immiscible liquid phases.
 
 \subsection{Lagrangian tracking and particle-interface interaction model} 
 
The motion of the particles is described by a set of ordinary differential equations for the particle velocity and position, which stem from the balance of the forces acting on the particles, which are assumed to be neutrally-buoyant ($\rho_p = \rho_f$, no effect of gravity) and smaller in size than the Kolmogorov length scale.
In this study,  we considered two force contributions: The drag force and the capillary force that is exerted on the particles when they interact with the interface, thus allowing for particle adhesion. With the above assumptions the Lagrangian equations of motion for the particles, in dimensionless vector form, read as

\begin{gather}
\frac{\partial \bf{x_p}}{dt}=\bf{u_p} ~~~, \hfill\\
 \notag \\
\frac{\partial \bf{u_p}}{\partial t}={\underbrace{\frac{\bf{u_{@p}}-\bf{u_p}}{St}(1+0.15Re_p^{0.687})}_{{\text{Drag force}}}}+{\underbrace{\frac{6A}{\rho_p /\rho_f} \frac{Re_{\tau}}{We} \frac{{\mathcal{D}}}{{d_p}^{3}}}_{{\text{Capillary force}}}} ~~~,
\label{LPT2}
\end{gather}

where $\bf{x_p}$ and $\bf{u}_p$ are the particle position and velocity, respectively; $\bf{u}_{@p}$ is the fluid velocity at particle position (obtained using a sixth-order Lagrange polynomials interpolation scheme);  $St=\tau_p/\tau_f$ is the Stokes number, ratio of the particle relaxation time $\tau_p=\rho_p d_p^2/18 \mu_f$ (with $d_p$ the particle diameter and $\mu_f$ the fluid dynamic viscosity) to the carrier fluid characteristic time $\tau_f=\nu_f /{u_{\tau}}^2$; $Re_{p}=|\bf{u_{@p}}-\bf{u_p}|d_p / \nu_f$ is the particle Reynolds number, introduced to correct the drag coefficient using the Schiller-Naumann correlation \cite{schiller1933drag} and ${\mathcal{D}}$ is the interaction distance between the center of mass of the particle and the nearest zero-level point on the fluid interface, which defines the range of action of the capillary force. 
%
%
The expression of the capillary force in Eq. (\ref{LPT2}) corresponds to the case of small spherical particles adsorbed at a fluid interface, and has been adopted in several previous studies to model particle particle-interface interactions \cite{sun2015three, gu2016direct, ettelaie2015detachment}.
%
Specifically, the dimensional expression of the force reads as
 \begin{equation}
  {\bf F}_{c}=
 \begin{cases}
           {\mathcal{A}} \pi\sigma {\mathcal{D}} {\bf n} \qquad \text{if} \ {\mathcal{D}} \le d_p \\
      0 \qquad \qquad \  ~~ \text{if} \ {\mathcal{D}} > d_p 
    \end{cases}
    \label{fipi}
 \end{equation} 

where ${\mathcal{A}}$ is a dimensionless parameter that characterises the magnitude of the capillary adhesion force (which incorporates the effect of the contact angle $\theta$ between the particle and the interface as well as the effect of the particle-to-drop size ratio) and
${\bf n}$ is the normal unit vector pointing from the particle center of mass to the zero-level set of $\phi$.
From Eq. (\ref{fipi}), it is clear that ${\bf F}_{c}$ reproduces the effect of a potential well centered at the interface that favours particle adhesion and attachment to the surface of the drop as soon as the particle touches the interface.
Albeit based on a mechanistic (rather than phsyics-based) model of the capillary force, Eq. (\ref{fipi}) represents the state of the art as far as particle-interaction models are concerned \cite{ gu2016direct}.
%
%
%

The value of the parameter ${\mathcal{A}}$ is chosen to satisfy the condition that the adsorption energy $E_{ads}=\pi \sigma r^2 (1-|\cos{\theta}|)^2$, corresponding to the difference between the energy of a particle fully displaced from the interface into the bulk phase and the energy of the particle settling at equilibrium at the interface, balances the desorption energy $E_{des}=\frac{1}{2} {\mathcal{A}} \pi \sigma r^2$,
corresponding to the energy required for particle detachment from the interface \cite{ettelaie2015detachment, gu2020fipi}.
Note that the expressions for $E_{ads}$ and $E_{des}$ are exact for an isolated, chemically homogeneous spherical particle on a flat surface \cite{gu2020fipi}. 
Assuming a contact angle $\theta=90^o$, this balance yields ${\mathcal{A}}=2$, which is the value used in our simulations.
Additional runs for different values of ${\mathcal{A}}$ (specifically: ${\mathcal{A}}=0.01$
and $0.1$) were also performed to assess the effect of a change in the magnitude of ${\bf F}_{c}$ on the capture process.
As far as the statistical quantities discussed in Sec. \ref{results} are concerned, no major effect was observed
(small quantitative modifications).

We remark that, in this study, only particles with tiny inertia are considered. For these particles, the capillary force is expected
to be dominant over the other hydrodynamic forces (drag, in particular) in close proximity of the interface, thus favoring particle
adhesion. However, this process may be significantly affected by the turbulence in the bulk of the carrier fluid, which tends to
deform continuously the interface and modify the topology of the flow structures with which the particles interact as they approach
the drop. Our aim is precisely to highlight the role played by these local flow structures and quantify their effect on
particle adhesion.

\subsection{Numerical method}
The governing equations (\ref{spfm}), (\ref{Mass}) and (\ref{NS}) are solved numerically using a pseudo-spectral method that transforms the field variables into wave space. Specifically, Fourier series are used to discrete the variables in the homogeneous directions (streamwise $x$ and spanwise $y$), while Chebychev polynomials are used in the wall-normal direction, $z$.
The Helmholtz-type equations so obtained are advanced in time using an implicit Crank-Nicolson scheme for the linear diffusive terms and an explicit two-step Adams-Bashforth scheme for the non-linear terms. Time-wise, the Cahn-Hilliard equation is discretized using an implicit Euler scheme, which allows damping of unphysical high frequency oscillation that may arise from the occurrence steep gradients in the phase field \cite{badalassi2003computation,yue2004diffuse}.
All unknowns (velocity and phase field) are Eulerian fields defined on the same Cartesian grid, which is uniformly spaced in $x$ and $y$
and suitably refined close to the wall along $z$ by means of Chebychev-Gauss-Lobatto points. Note that the Navier-Stokes equations are solved in their velocity-vorticity formulation and, therefore, are recast in a 4th order equation for the wall-normal component of the velocity and a 2nd order equation for the wall-normal component of the vorticity.

The Cahn-Hilliard equation is split into two second-order equations. Further details on the numerical method can be found in \cite{soligo2019coalescence}.

As far as boundary conditions are concerned, periodicity is imposed on all variables in $x$ and $y$, whereas a no-slip condition for velocity is enforced at the two walls, located at $z/h=\pm 1$

\begin{equation}
{\bf{u}}(z/h=\pm 1)=0 ~~~.
\label{BCWN}
\end{equation}

This condition yields the no-flux condition $\partial w / \partial z =0$ for the wall-normal velocity at $z/h=\pm 1$. The same condition is applied to the phase field

\begin{equation}
 \frac{\partial \phi}{\partial z} (z/h=\pm 1)=0 ~~~, \qquad 
 \frac{\partial^3 \phi}{\partial z ^3} (z/h=\pm 1)=0  ~~~.
 \label{BCPHI}
\end{equation}

These boundary conditions lead to the conservation of the integral of the phase field over time. \cite{yue2004diffuse}

\begin{equation}
\frac{\partial}{\partial t} \int_{\Omega} \phi d\Omega = 0 
\label{MassConserv}
\end{equation}

We remark here that the total mass of the carrier fluid and of the drops is conserved at all times, yet mass conservation of each phase is not guaranteed. To limit inter-phase mass leakage, we adopted the flux-corrected formulation proposed by \cite{li2016phase,zhang2017flux,soligo2019mass}.
In the simulations discussed here, this formulation limits mass leakage to roughly $5 \%$ of the drops during the initial time transient. At steady state, namely when the particle phase is also injected into the flow (see next paragraph), mass leakage vanishes.  

As far as the Lagrangian tracking is concerned, the particle equations of motion are integrated in time using an explicit Euler
scheme. Particles are injected into the flow once the surface area of the drops has reached a steady state: Particles are
initially placed at random locations within the volume occupied by the carrier fluid, namely in regions of the flow where $\phi = -1$ to avoid direct injection inside a drop, with initial velocity
${\bf u}_p ({\bf x}_p, t_{tr}=0)={\bf u}_{@p} ({\bf x}_{p}, t_{tr}=0)$, with $t_{tr}$ the particle tracking time. Interpolation of
flow variables, in particular fluid velocity components and phase field, at particle position is performed using 4th-order Lagrange polynomials. 

\subsection{Simulation setup}
The computational domain consists of a closed-channel configuration
with dimensions $ L_x \times L_y \times L_z = 4 \pi h \times 2 \pi h \times 2h $ ($ L_x^+ \times L_y^+ \times L_z^+ = 1885 \times 942.5  \times 300 $ in wall units). This domain is discretized using $N_x \times N_y \times N_z=512 \times 256 \times 257$ grid points,
which provide an extremely-well resolved turbulent flow field compared to the single-phase case (grid spacings are $\Delta x^+ = \Delta y^+ = 3.7$, $\Delta z^+_{wall}= 0.0113$ and $\Delta z^+_{center} = 1.84 \simeq \eta^+_{K,center} /2$, with $\eta^+_{K,center}$ the Kolmogorov length scale in the channel center. 
The flow is driven by a constant pressure gradient imposed along the streamwise direction, at shear Reynolds number $Re_{\tau}=150$.
We considered two different values of the surface tension corresponding to $We_L=0.75$ and $We_H=1.5$, respectively. These values match those commonly found in oil-water mixture \cite{than1988measurement}. In Sec. \ref{results}, the simulation results are discussed with reference to $We_L$, which corresponds to less deformable drops. However, we remark here that the effect of the Weber number observed in our simulations is limited to minor quantitative modification so the statistical quantities examined.
For the phase field, the value of the Cahn number has been set to ensure that there are at least five grids points across the interfacial layer to resolve accurately all the gradients occurring there \cite{soligo2019coalescence, soligo2020effect}. This condition yields $Ch=0.02$.
The P$\acute{\mathrm{e}}$clet number has been set according to the scaling  $Pe_{\phi}=1/Ch=50$ proposed by
\cite{magaletti2013sharp} to achieve the convergence to the sharp interface limit.

At the beginning of the simulations, the phase field was initialized to generate a regular array of 256 spherical drops with normalised diameter $d/h=0.2$ (corresponding to $d^+=60$ in wall units) that are injected in a fully-developed turbulent flow. Particles were released into the flow only after the surface area of the drops had reached a steady state, which results from a balance between coalescence and breakup events for the range of Weber numbers considered in this study. A total of five sets of ${\mathcal{N}}_p=10^6$ particles at varying Stokes number were tracked: Tracer particles ($St=0$), which are used as markers to sample all flow regions in the carrier fluid domain, and particles with $St=0.1$, 0.2, 0.4 and $0.8$, corresponding to particle diameters much smaller than the drop diameter - $d_p / d \simeq {\mathcal O} (10^{-2})$ at least.

Since the focus of the present study is on particle capture by turbulence, the feedback of particles on the flow field is not considered (one-way coupling simulation). Indeed, particles are characterized by low values of inertia and exhibit a weak tendency to cluster: This implies that their spatial distribution within the carrier fluid domain remains dilute over the entire simulation. Also, particle-particle collisions are not accounted for: these are assumed to be negligible prior to particle adhesion to the drop interface.

\section{Results and discussion}
\label{results}

In this Section, we will first characterize the process of particle capture at the drop interface, focusing in particular on the
topology of the flow structures that drive particle adhesion. Unless otherwise stated, the statistics presented in the following refer
to a steady-state condition for the surface area of the drop swarm. Then, we will discuss the macroscopic outcome of this process,
the time accumulation of particles on the interface, and propose a simple model to estimate the rate at which this accumulation takes place.

\subsection{Particle capture and flow topology}

\begin{figure}[t]
 \vspace{-0.5cm}
%
%
%
	\includegraphics[width=105mm, keepaspectratio, angle=0]{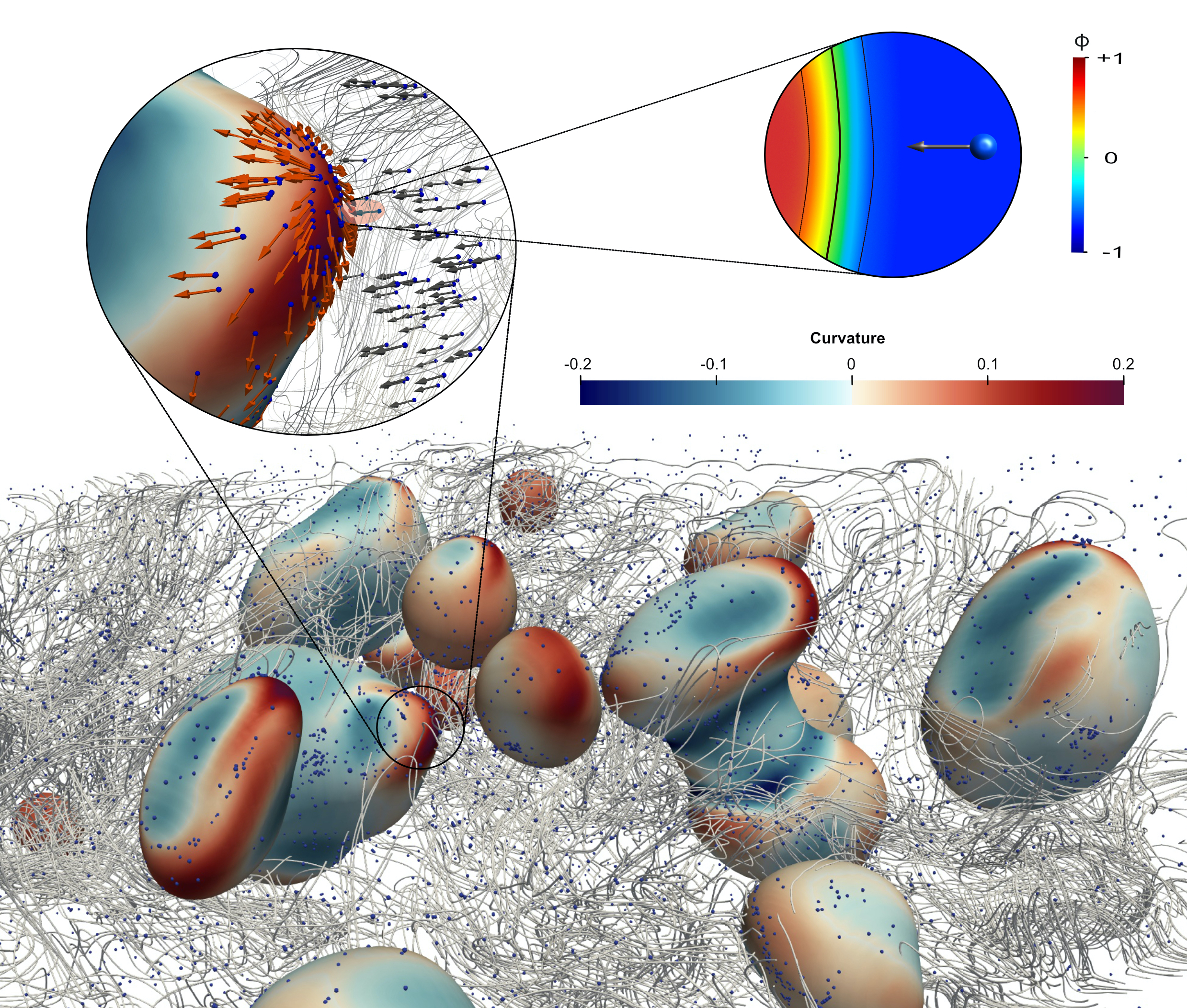}
			\caption[]{Qualitative rendering of the flow configuration. drops are coloured by the local curvature of the interface, the flow field is rendered by the fluid streaklines and particles are visualized as blue spheres. The insets provide close-up views of one particle capture event (inset $I$) and of one interface-approaching particle in isolation (inset $II$), respectively. Grey arrows represent the particle velocity magnitude and render the motion of particles that move towards the interface. Red arrows represent the interfacial stress sampled by the particles at the time of adhesion. The colormap in the top inset shows the spatial distribution of the phase indicator $\Phi$: The interface is located at $\Phi=0$ (thick black line). The thin black lines represent the fluid layer within which $\Phi$ transitions from
			$\Phi=-1$ (fluid) to $\Phi=+1$ (drop).
			}
\label{flow-snapshot}

\vspace{-16.8cm}
\hspace{-8.6cm}
$I$

\vspace{-1.3cm}
\hspace{2.9cm}
$II$

\vspace{17.0cm}
\end{figure}

A qualitative rendering of the instantaneous flow field is provided in Fig. \ref{flow-snapshot}, where a close-up view of one
capture event is also shown. The carrier phase is rendered by means of the fluid streaklines. drops are visualised by the
$\Phi=0$ iso-surface and are coloured by the local curvature of the surface (concave areas with high negative curvature are
shown in blue, convex areas with high positive curvature are shown in red).
Particles are represented as blue dots, with size equal to the particle diameter ($St=0.1$ particles are considered here).
Note that captured particles tend to form
filamentary clusters, which result from the action of the capillary force ${\bf F}_c$. The mechanisms that lead to the formation
of these clusters and their topological characterization is beyond the scope of this paper, and will be the subject of an independent study
focusing on particle dynamics after capture. It suffices to say here that the formation of neat particle filaments is favoured by the
neglect of inter-particle collisions, which are expected to smear out densely-concentrated clusters.
  
Two close-up views are provided: One
(marked as $I$) shows a near-drop region of the flow populated by a swarm of particles that is being pushed toward the
drop by the carrier fluid, the other (marked as $II$) shows one isolated particle approaching the interface with the phase field
distribution in background.
At this time of the simulation, the total surface area of the drops has reached a statistically-steady state that results from
a balance of (now rare) coalescence and breakup events. drop deformation induced by turbulence is apparent and is associated
to a non-uniform distribution of the curvature, which can be computed starting from the phase field as
\begin{equation}
\kappa=-\nabla \cdot \left ( \dfrac{\nabla \Phi}{| \nabla \Phi |} \right ) =
-\dfrac{\nabla^2 \Phi}{| \nabla \Phi |}+\dfrac{1}{| \nabla \Phi |^2} \nabla \Phi \cdot \nabla (| \nabla \Phi | ) ~~~.
\label{curvature}
\end{equation}
In addition, the local unit vector {\bf n} normal to each level-set curve is obtained as
\begin{equation}
{\bf n}=- \dfrac{\nabla \Phi}{| \nabla \Phi |} ~~~,
\label{normal}
\end{equation}
where equations (\ref{curvature}) and (\ref{normal}) are valid only if $\Phi$ iso-surfaces are parallel to each other. This property is
conserved when advecting $\Phi$ through the Cahn-Hilliard equation using the $Pe \propto Ch^{-1}$ scaling \cite{magaletti2013sharp}.

Focusing on inset $I$, we observe that particles tend to approach the drop and adhere to its surface in a convex region of the
interface where curvature $\kappa$ reached a local peak. In this region, the flow is impinging on the drop surface and the tangential shear stress is directed from the high-positive curvature region towards the neighbouring, high-negative curvature regions.
This anticipates that captured particles, while subject to the action of tangential stresses, will be driven toward such regions as long as they remain attached to the interface. 
 
%
%

Fig. \ref{flow-snapshot} confirms the physical intuition that particles are brought in close proximity of the drop by coherent fluid motions that interact
with the compliant drop surface. This interaction gives rise to highly non-uniform curvature and shear stress distributions.  
In order to examine these fluid motions in more detail, we consider first the two-dimensional fluid velocity divergence at the interface
of the drop, referred to as surface divergence in the following.
The surface divergence is defined as
\begin{equation}
\nabla_{2D}=\bf{n} \cdot \nabla \times (\bf{n} \times \bf{u}) 
\label{Div2D}
\end{equation}
According to this definition, particles captured at the surface probe a compressible two-dimensional
system where regions of local flow expansion, generated by impinging fluid motions, are characterized
by $\nabla_{2D} > 0$  and regions of local compression, generated by outward fluid motions, are characterized
by $\nabla_{2D}  < 0$.

\begin{figure}
 \vspace{0.5cm}
\centering	
	\includegraphics[width=125mm, keepaspectratio, angle=0]{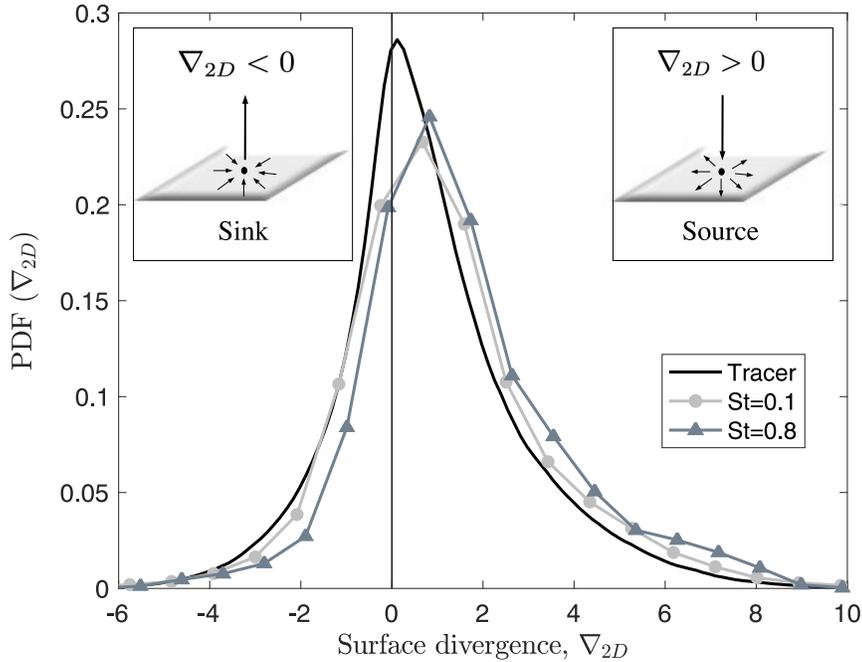}	
			\caption[]{PDF of the 2D surface divergence, $\nabla_{2D}$, seen by the particles when they get captured by the interface.
			Regions of local flow expansion (velocity sources) are characterised by $\nabla_{2D} > 0$, regions of local compression
			(velocity sinks) are characterized by $\nabla_{2D}  < 0$.
			Symbols refer to simulation results (circles: $St=0.1$, triangles: $St=0.8$), whereas the solid line refers to the PDF 
			computed for tracer particles uniformly distributed over the entire interface of each drop. The PDF was computed
			starting at time $t^+ \simeq 1000$ after particle injection and over a subsequent time interval $\Delta t^+=2000$.
			}
\label{surfdiv}
\end{figure}

In Fig. \ref{surfdiv}, we show the probability distribution function (PDF) of the surface divergence computed at the
position occupied by the particles when they get captured by the interface. This position is evaluated at the time the
particle touches the interface, namely when the particle center is less than one radius away from the nearest zero-level
point on the interface. To allow comparison among the different particle sets, we considered a reference
distance equal to the radius of the largest particles ($St=0.8$), which is equal to about one tenth of the interface
thickness and thus corresponds to a phase field $\Phi = -0.71$.
The PDFs for the $St=0.1$ and $St=0.8$ are shown, and compared to the PDF obtained for the case of inertialess tracers
uniformly distributed over the $\Phi = -0.71$ iso-surface.
We remark that, in our simulations, this is also distance within which the capillary force ${\bf F}_c$
starts acting on the particle. Therefore, the PDFs shown in Fig. \ref{surfdiv} is not affected by the model used for ${\bf F}_c$
in the equation of particle motion.

Fig. \ref{surfdiv} shows that, in the case of the tracers, the PDF exhibits a clear peak at $\nabla_{2D} = 0$ but is also negatively
skewed. This indicates that fluid motions directed
towards the drop occupy a wider surface area as compared to fluid motions directed away from the drop. The effect can
be ascribed to the deformability of the interface, which is able to respond and adapt {\it elastically} to impinging flow events. 
In the case of particles with tiny inertia, the PDF shifts towards higher positive values of $\nabla_{2D}$: The peak is now located
at $\nabla_{2D} \simeq 1$, and inertia appears to play a negligible role for the range of Stokes numbers considered in the study.
Overall, Fig. \ref{surfdiv} corroborates the observation that particles tend to sample preferentially regions of local flow expansion
as they attach to the drop. This provides already a first indication about the topological features of the flow near the interface.

To provide additional information about these features, we examine next the flow topologies that are sampled by the particles just before being captured. A flow topology analysis near deformable drops has been carried out recently by
\cite{dodd2019small} for the case of decaying isotropic turbulence. Following the classification proposed by
\cite{perry1987description} (to which the reader is referred to
for a detailed discussion of the flow topologies in three-dimensional flow fields), these Authors showed that there is a shift from high-enstrophy/low-dissipation structures favoured outside the near-surface viscous layer to low-enstrophy/high-dissipation structures favoured inside the viscous region and, eventually, to boundary-layer like and vortex-sheet flow topologies at the surface. In close proximity of the surface, the observables examined to characterized the topological structures (the invariants of the velocity-gradients, rate-of-strain and rate-of-rotation tensors) exhibit statistical features that are very similar to those reported inside the viscous sublayer of wall-bounded turbulence \cite{dodd2019small, picciotto2005characterization}.
The analysis we propose is thus justified by the expectation that the final particle capture rate will result from particle interaction with all these topological structures. 
To infer the local flow structures sampled by the particles, we use standard observables that are related to the invariants of the velocity gradients tensor ${\mathcal{A}} = [u_{i,j}]$ \cite{chong1990general,blackburn1996topology}
%
%
\begin{equation}
P=- tr [{\mathcal{A}}] ~~~,
\end{equation}
\begin{equation}
\label{Q}
Q=\frac{1}{2} (P^2 - {\textit{tr}} [ {\mathcal{A}}^2 ] ) ~~~,
\end{equation}
\begin{equation}
R=- det [ {\mathcal{A}} ] ~~~.
\end{equation}
Note that, for incompressible flow, $P=0$ and the second invariant can be expressed simply as $Q=-\frac{1}{2}(S:S + \Omega:\Omega)=-\frac{1}{2}(S^2 + \Omega^2)$, where $S= \frac{1}{2}(u_{i,j}+u_{j,i})$ and $\Omega= \frac{1}{2}(u_{i,j}-u_{j,i})$ and  the symmetric and antisymmetric components of ${\mathcal{A}}$, respectively. In this case, $Q$ represents the local balance between vorticity (related to $\Omega$) and strain rate (related to $S$).
%
Thus, a fluid point characterized by
positive values of $Q$ indicates the presence of high vorticity, whereas for negative values of $Q$ the local flow is dominated by straining motions \cite{jeong1995identification}. Based on $Q$, the following topology parameter can be defined  \cite{de2019effect,rosti2019numerical,de2017viscoelastic,soligo2020effect}
\begin{equation}
\label{topolparam}
{\mathcal{Q}}=\frac{S^2 - \Omega^2 }{S^2+\Omega^2} ~~~.
\end{equation}
Based on this definition, ${\mathcal{Q}}=1$ corresponds to purely elongational flow ($\Omega=0$), ${\mathcal{Q}}=0$ corresponds to shear flow and ${\mathcal{Q}}=-1$ corresponds to purely rotational flow  ($S=0$) \cite{soligo2020effect}.
The topology parameter has been used recently to examine the effect of a compliant interface on the flow field in different regions of the flow domain in two-phase systems \cite{de2019effect,rosti2019numerical,soligo2020effect}.

\begin{figure}[t]
 \hspace{-0.0cm}
%
%
%
%
	~~~~ \includegraphics[width=85mm, keepaspectratio, angle=0]{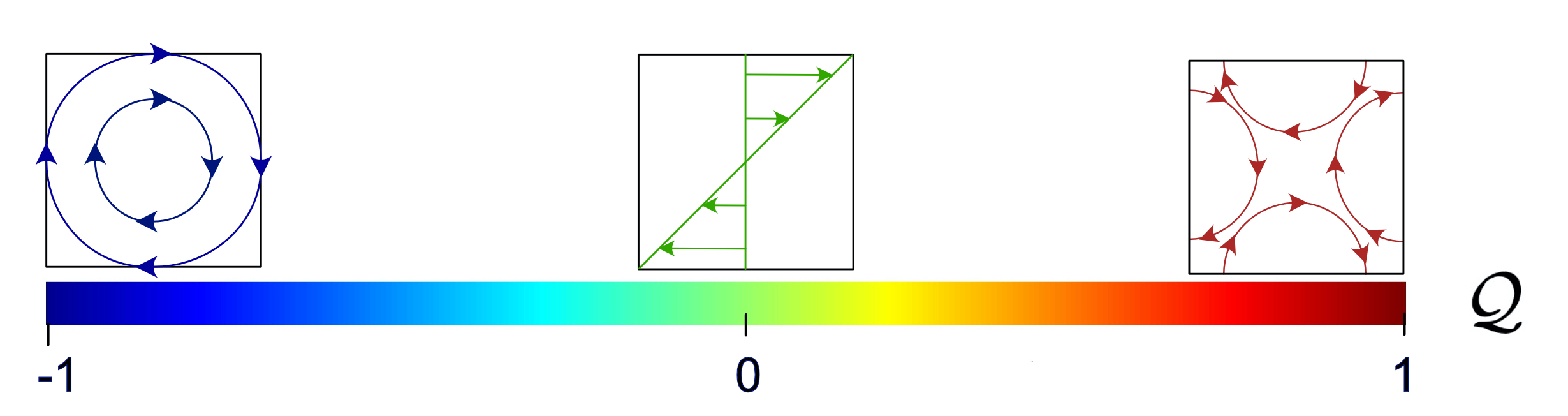}	
	\includegraphics[width=135mm, keepaspectratio, angle=0]{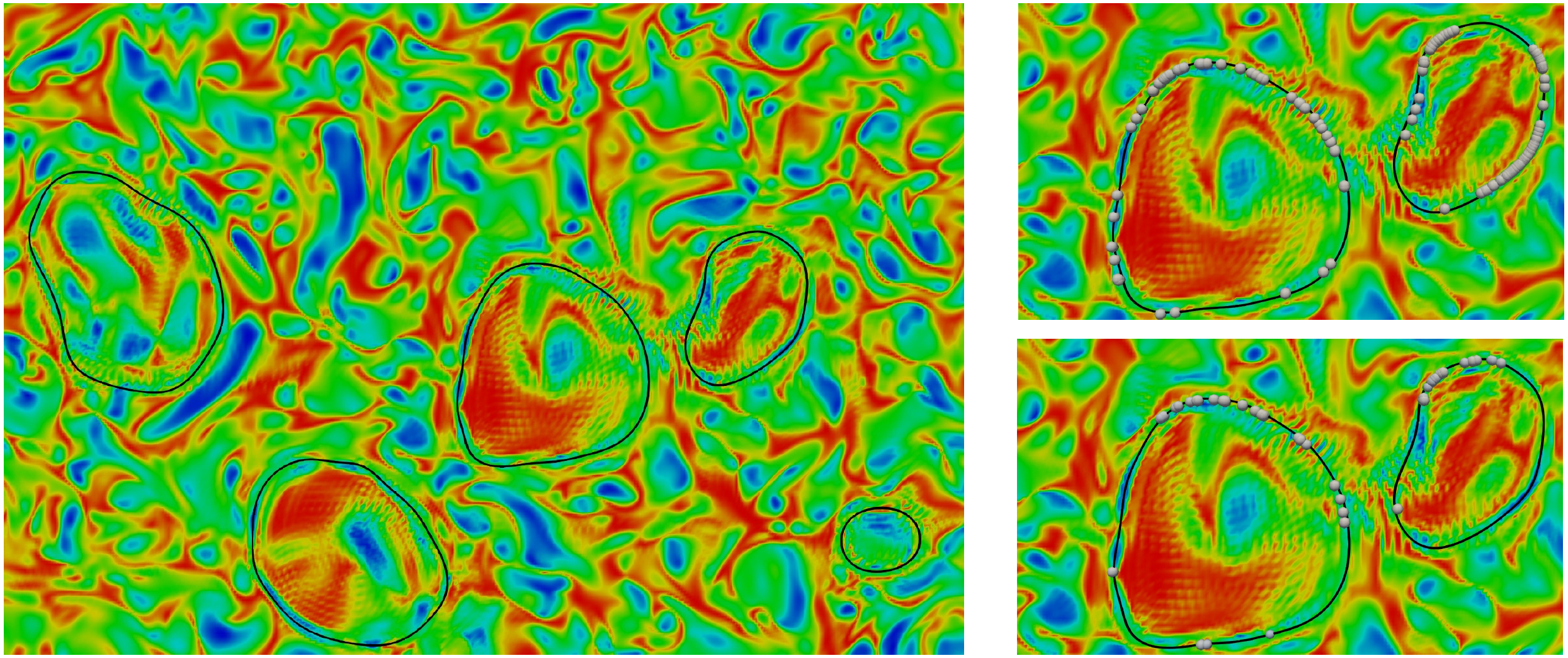}
			\caption[]{(a): Flow topology parameter, ${\mathcal{Q}}$, on the channel midplane ($x^+ - y^+$ plane); black solid lines identify the position of the drop interface (isolevel $\phi=0$). (b)-(c): Close-up view of the distribution of captured particles on the interface of the drop pair boxed in panel (a). Insets: (b) $St=0.1$, (c) $St=0.8$.
			}
\label{topolparqual}

\vspace{-9.4cm}
\hspace{-0.05cm}
(a)
\hspace{13.3cm}
(b)

\vspace{2.1cm}
\hspace{14.0cm}
(c)

\vspace{-1.64cm}

\begin{tikzpicture}
\hspace{-1.5cm}
\draw[black,thick] (0,0) rectangle (3.8,2.2);
\end{tikzpicture}

\vspace{0.5cm}
\hspace{-14.0cm}
$\Bigg\uparrow$

\vspace{-0.2cm}
\hspace{-12.8cm}
o $\xrightarrow{~~~~~~~~}$ x

\vspace{-2.0cm}
\hspace{-14.5cm}
y

\vspace{4.7cm}
\end{figure}
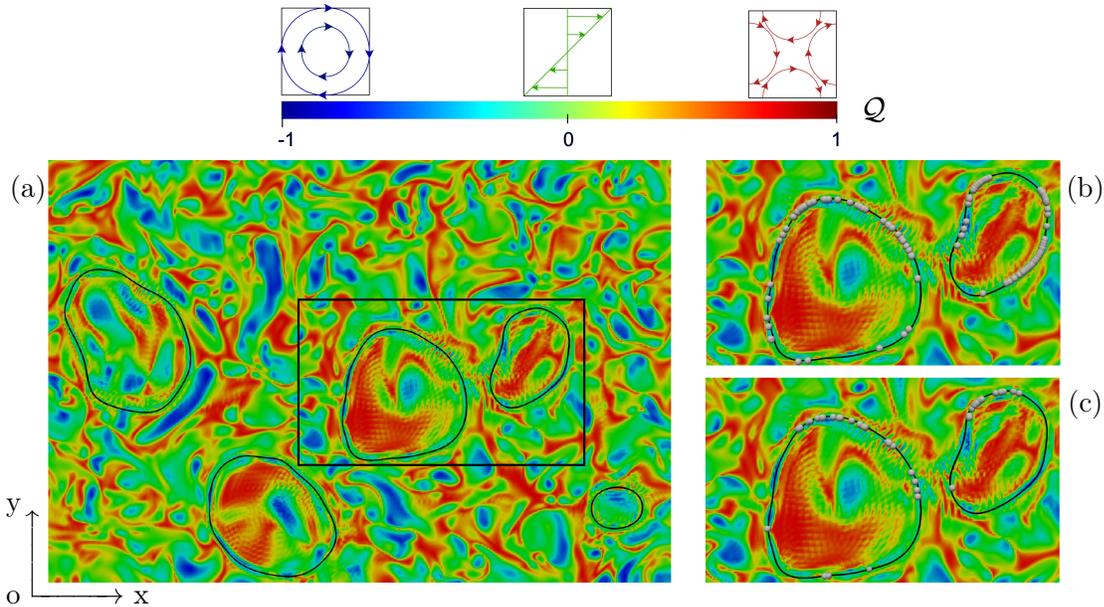

Fig. \ref{topolparqual} shows the instantaneous spatial distribution of ${\mathcal{Q}}$ in the wall-parallel $x^+ - y^+$ plane at the center of the channel. The interface of the drops is represented by the black solid lines. Panel (a) refers to the entire $x^+ - y^+$ plane, whereas the two insets show, for $St=0.1$ and $St=0.8$ respectively, a close-up view of particle distribution along the surface of the drop pair highlighted in panel (a). The presence of the interface has a clear influence on the local flow behavior. The carrier phase appears to be
characterized by large areas of shear flow (in green, corresponding to values of ${\mathcal{Q}}$ close to zero), and smaller fragmented
regions of rotational flow (in blue, corresponding to values of ${\mathcal{Q}}$ close to -1) and elongational flow (in red, corresponding to values of ${\mathcal{Q}}$ close to +1). The flow inside the drops, on the other hand, is most often characterized by the predominance
of both shear and elongational flow regions, as also noted by
\cite{soligo2020effect}. 
The insets show that small changes of particle inertia are sufficient to modify the spatial distribution of the captured particles over the interface.
Note that, at the Weber number values considered in this study, only a small number of drops is found at steady state: Therefore, the drop size is large enough to
minimise the internal flow confinement effects that are observed at higher Weber number \cite{soligo2020effect}. 

\begin{figure}[t]
 \vspace{0.5cm}
\centering	
	\includegraphics[width=125mm, keepaspectratio, angle=0]{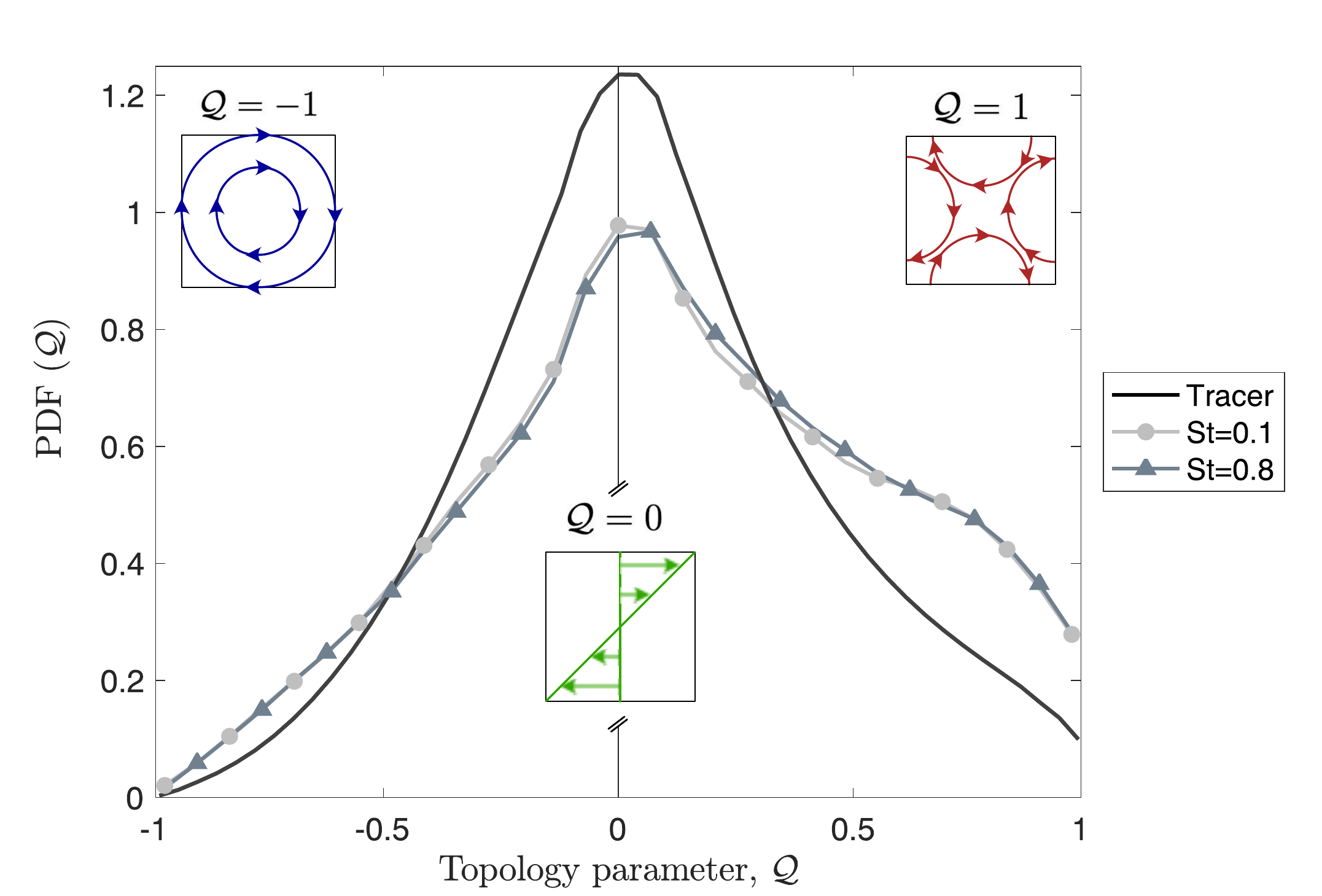}
			\caption[]{PDF of the topology parameter, ${\mathcal{Q}}$, seen by the particles when they get captured by the interface. Lines and symbols are as in Fig. \ref{surfdiv}. The PDF was computed
			starting at time $t^+ \simeq 1000$ after particle injection and over a subsequent time interval $\Delta t^+=2000$.
			}
\label{pdftop}
\end{figure}

It is not so easy to conclude something about the flow behavior very close to the interface just by visual inspection of Fig. \ref{topolparqual}.
To this aim, in Fig. \ref{pdftop} we show the PDF of ${\mathcal{Q}}$ seen by the particles at the
time they touch the interface and get captured. As done for Fig. \ref{surfdiv}, ${\mathcal{Q}}$ is thus evaluated when the phase
field value interpolated at particle position is $\Phi = -0.71$, namely at the edge of the capillary force range: This prevents any
effect of this force on the motion of the particles in their final stretch to the interface. Lines and symbols are as in Fig. \ref{surfdiv}.
For the case of inertialess tracers, the PDF is slightly asymmetric and negatively skewed, indicating that elongational flow events
(${\mathcal{Q}}>0$) are slightly more likely than rotational flow events (${\mathcal{Q}}<0$). Interestingly, a small amount of
particle inertia is sufficient to produce a significant quantitative change in the shape of the PDF: Asymmetry is increased and
the likelyhood of particles sampling shear-dominated flow events decreases in favour of elongation-dominated events. As particles
reach the very-near interface region, the interplay between the impinging fluid motions that are transporting the particles and
the blockage effect of the interface generates stronger tangential stresses, which in turn generate localized elongational flows
similar to that highlighted in inset $I$ of Fig. \ref{flow-snapshot}. Strong rotational flow events also become slightly more likely,
but this seems to be a minor effect.
 
\begin{figure}[t]
 \vspace{-1.5cm}
 \hspace{0.66cm}
\centering	
	\includegraphics[width=59.7mm,height=60.2mm, angle=0]{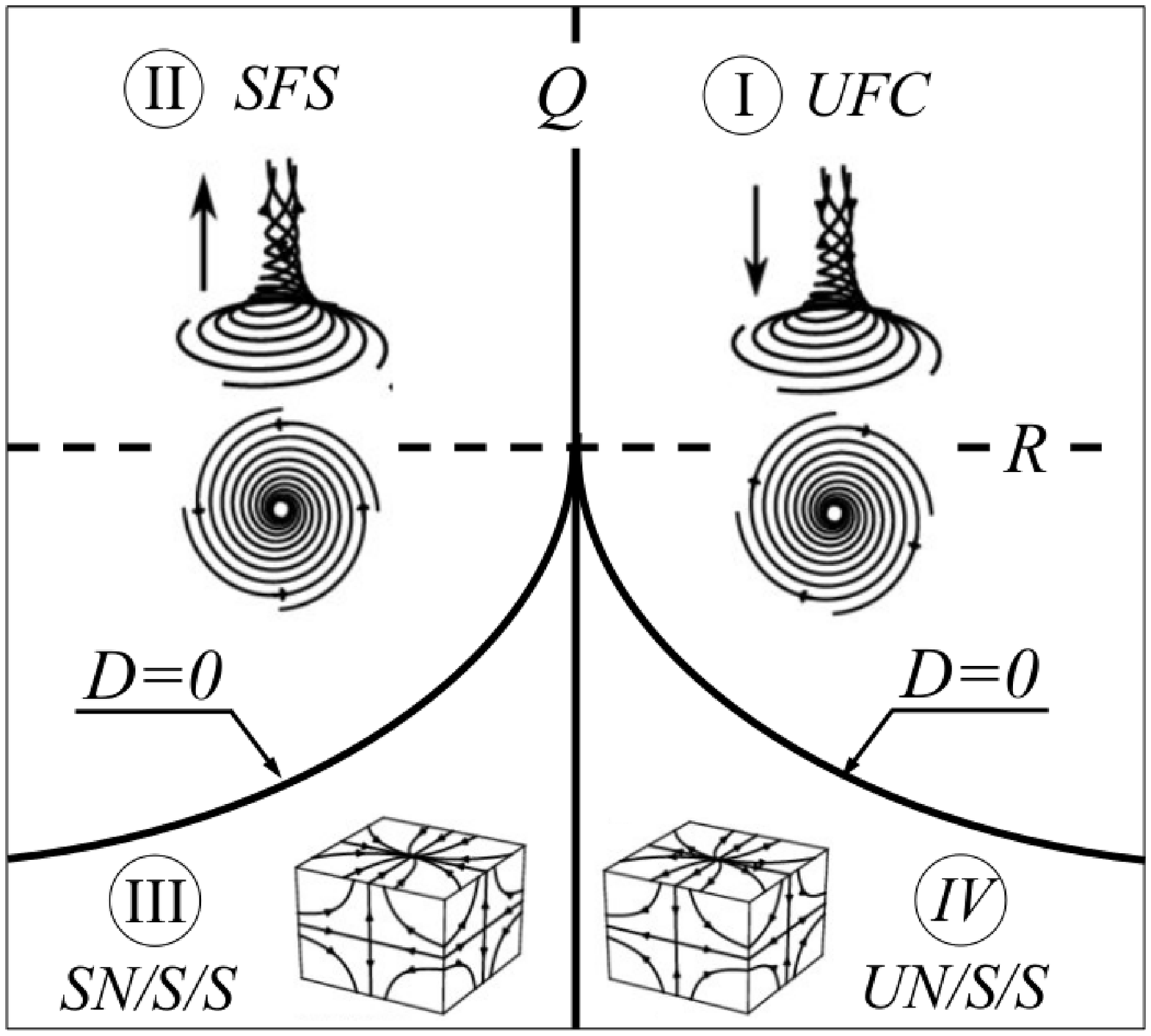}
	\hspace{-0.24cm}
	\includegraphics[width=81.5mm, keepaspectratio, angle=0]{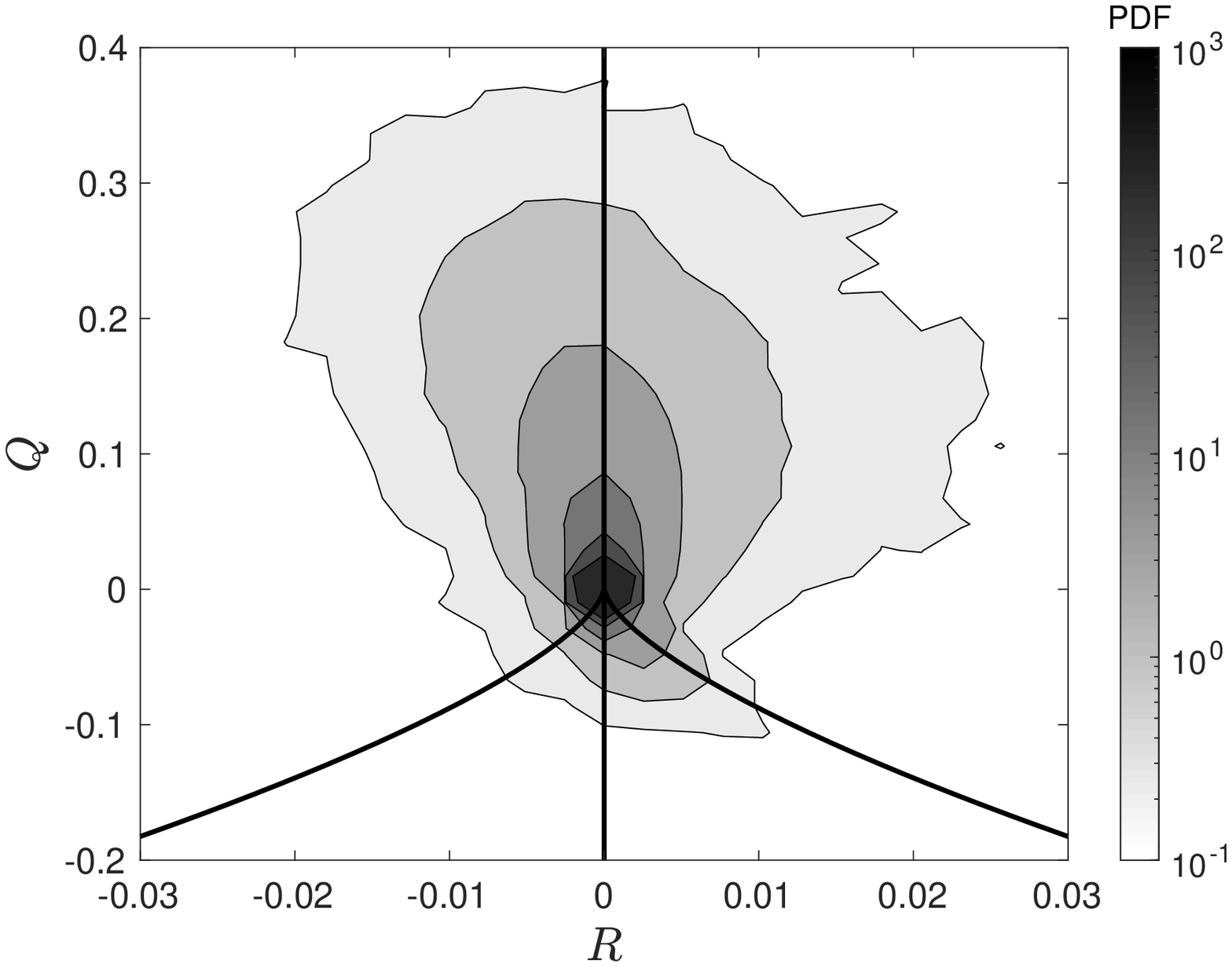}
		\includegraphics[width=81.5mm, keepaspectratio, angle=0]{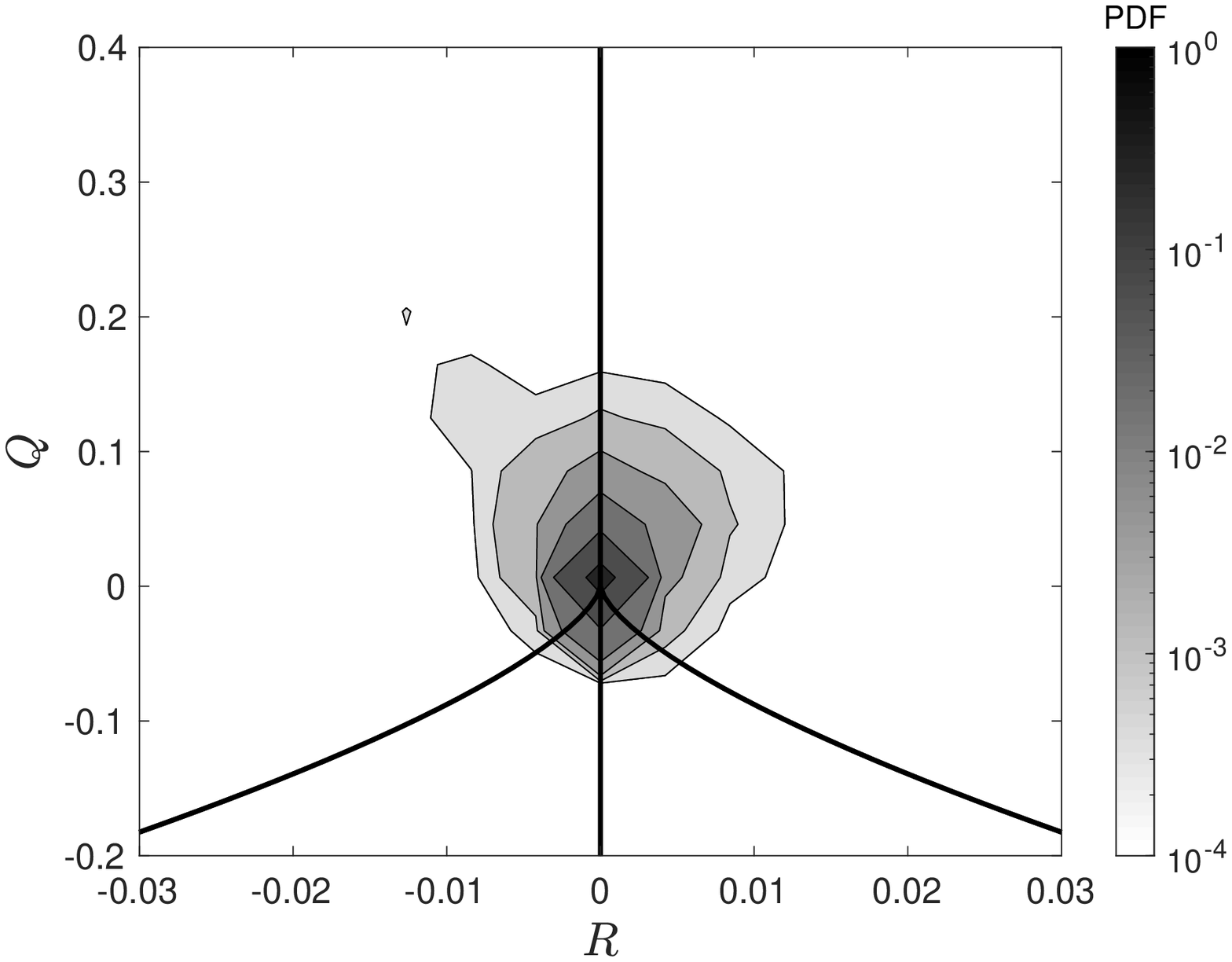}
	\hspace{-1.6cm}
	\includegraphics[width=81.5mm, keepaspectratio, angle=0]{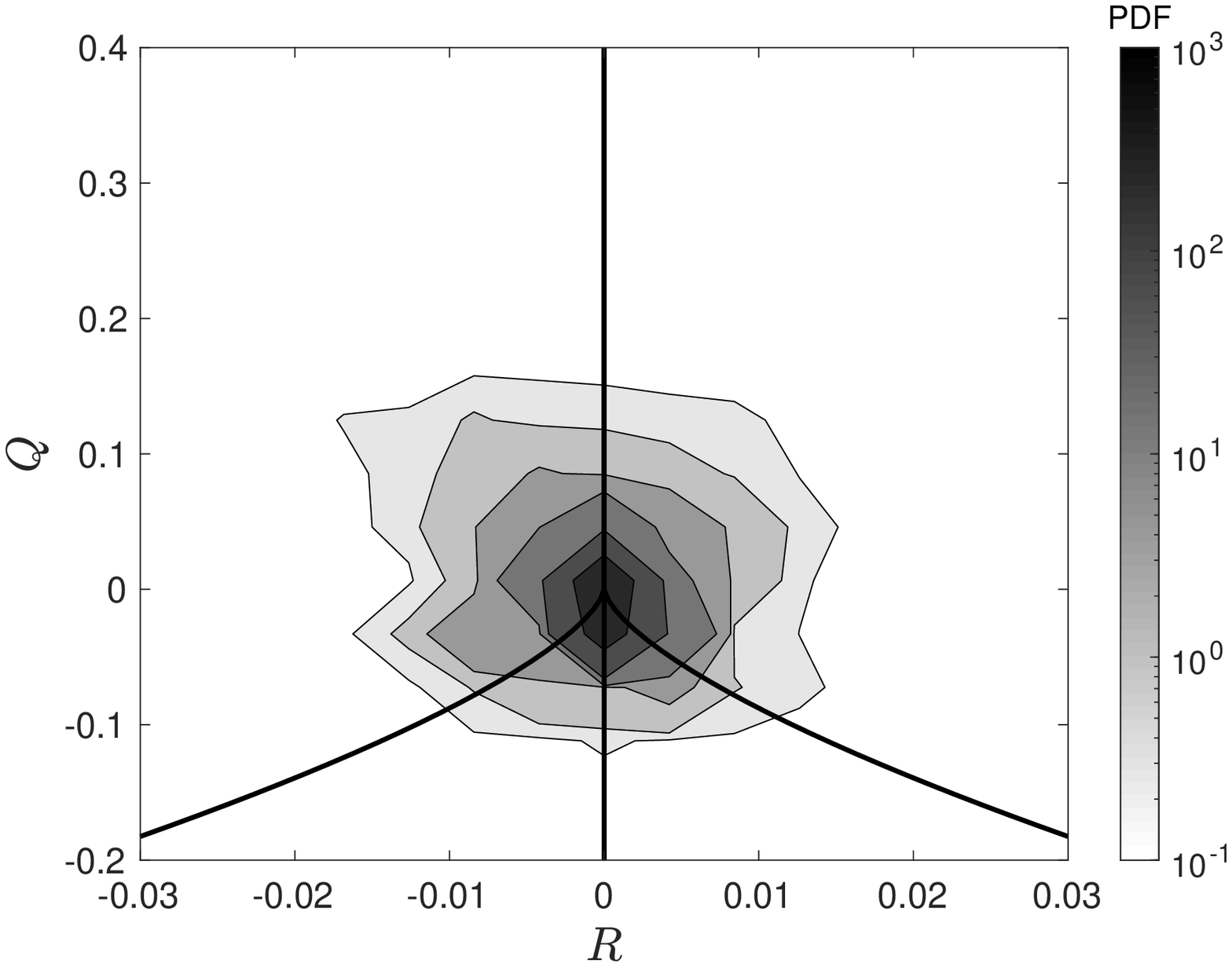}
			\caption[]{
			Panel (a): Incompressible flow critical point topologies according to the classification
scheme by
\cite{chong1990general}. Panels (b)-(d): Joint PDF of $Q$, $R$ conditionally sampled for fluid at grid points over
the entire interface of the drops (b), at the position of the
$St=0.1$ particle when they get captured (c), and the position of the $St=0.8$ particles when they get captured.
			}
\label{goccia}

\vspace{-16.2cm}
\hspace{-2.6cm}
(a)

\vspace{-0.7cm}
\hspace{11.0cm}
(b)

\vspace{5.4cm}
\hspace{-2.6cm}
(c)

\vspace{-0.75cm}
\hspace{10.8cm}
(d)

\vspace{-2.25cm}
\hspace{-6.8cm}
\begin{tikzpicture}
\node[text=black,font=\fontsize{1}{1}\selectfont]{UFC=Unstable Focus Compressing};
\end{tikzpicture}

\vspace{-0.52cm}
\hspace{-6.8cm}
\begin{tikzpicture}
\node[text=black,font=\fontsize{1}{1}\selectfont]{SFS=Stable Focus Stretching};
\end{tikzpicture}

\vspace{-0.52cm}
\hspace{-6.8cm}
\begin{tikzpicture}
\node[text=black,font=\fontsize{1}{1}\selectfont]{SN/S/S = Stable Node/Saddle/Saddle};
\end{tikzpicture}

\vspace{-0.52cm}
\hspace{-6.8cm}
\begin{tikzpicture}
\node[text=black,font=\fontsize{1}{1}\selectfont]{UN/S/S = Unstable Node/Saddle/Saddle};
\end{tikzpicture}

\vspace{10.5cm}

\end{figure}

To conclude the analysis of the flow events that drive particle capture, we examine their topological features by discussing the
joint PDF of the second and third invariants of the velocity gradient tensor, $Q$ and $R$. These invariants are computed at the
Eulerian grid points and then interpolated at the instantaneous position of particle capture using fourth-order Lagrange polynomials: Near the interface,
a one-sided version of the scheme is used to avoid mixing drop- and carrier-fluid velocities \cite{dodd2019small}.
The time window considered to compute the invariants covers the last $400$ viscous units of the simulations.
The conditioned joint PDFs so obtained are shown in Fig. \ref{goccia}. 
For clarity of presentation, in panel (a) of this figure we show first a compact classification of all incompressible flow topologies in the
($Q$,$R$)-plane \cite{perry1987description,blackburn1996topology}. This classification also involves the discriminant
$D=\frac{27}{4} R^2 + Q^3$ of the velocity gradient tensor: If $D>0$ then the tensor has one real and two complex-conjugate
eigenvalues indicating prevalence of enstrophy in the flow, if $D<0$ then the tensor has three real, distinct eigenvalues indicating prevalence of dissipation in the flow, if $D=0$ then the tensor has three real eigenvalues, two oh which are equal \cite{dodd2019small}.
Based on the sign of $D$ and $R$, four topological regions can be identified: When $D>0$ and $R>0$ (region $I$), the flow is characterized by predominance of vortex compression over vortex stretching and the opposite is true when $D>0$ and $R<0$
(region $II$); when $D<0$ and $R<0$ (region $III$), the flow is connected to diverging fluid trajectories while being connected to
converging trajectories when $D<0$ and $R>0$ (region $IV$). Using the terminology adopted by
\cite{chong1990general},
topologies falling in region $I$ are called stable focus/compressing while those falling in region $II$ are called unstable focus/stretching;
topologies falling in region $III$ are called stable node/saddle/saddle while those falling in region $IV$ are called unstable node/saddle/saddle.
Further critical points can be identified along the $Q$-axis and the $D=0$ line, but their characterization is beyond the scope of this study.
%

Fig. \ref{goccia}(b) shows the joint PDF conditioned at the position of uniformly-distributed tracers. This PDF is characterized
by the same teardrop shape that is typically observed in wall-bounded flows, and particularly in the viscous sublayer region.
There is also evidence of events clustered around $Q=0$ and $R=0$, which are indicative of boundary-layer-like flow topologies
 \cite{dodd2019small}.
This confirms that, at least from a qualitative viewpoint, there are similarities between the flow field near a compliant interface
and the flow field near a solid wall.
The most probable flow topologies are those falling in regions $I$ and $II$, as also shown in Table $I$:
These topologies represent vortical motions
that contribute to the production of enstrophy via vortex compression or stretching, respectively.
When we consider the joint PDF conditioned at the particle position, there is a clear change of shape and the so-called Vieillefosse tail in the lower right quadrant disappears. Comparing also the percent values reported in Table $I$, we observe a non-trivial effect of the Stokes number. For the $St=0.1$ particles, we find an increased probability associated to unstable focus/compressing topologies and a nearly equivalent decrease of the probability associated to stable focus/stretching topologies with respect to tracers (the sum of the two probabilities being equal to $77 \%$). Percent values associated to node/saddle/saddle topologies are almost unchanged, and sum up to just $23 \%$ of the occurrences. This indicates that $St=0.1$ particles sample preferentially high-enstrophy fluid motions that
contribute to enstrophy production via vortex compression more often than via vortex stretching.
Fluid motions characterized by high strain and high dissipation are avoided by these particles.
For the $St=0.8$ particles, probabilities are more evenly distributed with a significant decrease for the case of unstable focus/compressing topologies (with respect to tracers but also $St=0.1$ particles) and a significant increase for the case of node/saddle/saddle topologies,
particularly unstable ones. These are regions with large negative values of $Q$ and represent sites of high dissipation that $St=0.8$ particles apparently sample just before adhesion.


\begin{table}[t]
\begin{center}
\begin{tabular}{ |c||c|c|c|c| }
\hline
Quadrant    &  $I$  &  $II$   &  $III$  &  $IV$   \\ 
 \vspace{-1.0cm}
                   & & & & \\
{\tiny{Topology}}   & {\tiny{Unstable focus/}}  & {\tiny{Stable focus/}}  &   {\tiny{Stable node/}}  &  {\tiny{Unstable node/}}  \\ \vspace{-1.2cm}
                   & & & & \\
                   & {\tiny{compressing}}  & {\tiny{stretching}}  &   {\tiny{saddle/saddle}}  &  {\tiny{saddle/saddle}}  \\ 
\hline \hline
 Tracers & $35 \%$ & $45 \%$ &$7 \%$ &$13 \%$  \\ 
\hline 
 $St=0.1$ &$42.5 \%$ & $34.5 \%$ &$9 \%$ &$14 \%$  \\ 
 \hline 
 $St=0.8$ &$25 \%$ &$33 \%$ &$15 \%$ & $27 \%$  \\  
 \hline
\end{tabular}

\end{center}
\caption{Probabilities representing the tendency of captured particles to sample the different incompressible topologies near the interface of the drops. Probabilities are averaged over the last $400$ viscous time units of the simulations.}
\end{table}

\section{Particle capture rate}

The phenomenology of particle capture by the drop is as follows: A flow event, roughly described as a jet, transports the particles towards the interface; near the interface, the jet deflects and particles that are close enough are captured by the interfacial forces.
This phenomenology is by no means different than that controlling particle deposition
at a solid wall \cite{fried,cleaver,c&h,soldati96,gravity2007,ijmf2009}.
Starting from this similarity, in this section we
propose a simple mechanistic model that can be used to obtain a reliable prediction of the capture rate
and, at the same time, can easily be implemented in industrially-oriented CFD codes.

In general, there are three main deposition mechanisms that may act simultaneously: diffusion, impaction
and interception. However, at fluid velocities typical of scrubbing devices and for micron-sized particles like
those considered in the present study, impaction is known to be the dominant capturing mechanism \cite{Kim2001}.
In this case, the classical deposition models by \cite{fried,c&h,soldati96} assume that the deposition rate of non-interacting
particles is proportional to the ratio between the mass flux of particles at the deposition surface, $J$, and the mean
bulk concentration of particles, $C$. Through the definition of a suitable constant of proportionality, usually referred to as the deposition
coefficient $k_d$, the following turbulent transport equation holds
\begin{equation}
J = k_d C
\end{equation}
Given the initial number $N_0$ of particles released
in the carrier fluid sub-domain, $J$ and $C$ can be discretized as follows
\begin{equation}
\label{j}
J = \frac{1}{A} \cdot \frac{dN_{c}(t)}{dt} ~~~,
\end{equation}
\begin{equation}
\label{c}
C = \frac{N_0 - N_{c}(t)}{V} ~~~,
\end{equation}
where $N_{c}(t)$ is the number of particles captured by the interface at time
$t$, $A$ is the total surface area of the drops and
$V$ is the volume occupied by the carrier fluid. These definitions yield
\begin{equation}
\label{dndt}
\frac{dN_{c}(t)}{dt} = k_d [ N_0 - N_{c}(t) ] \frac{A}{V} ~~~.
\end{equation}
Once $k_d$ is known, Eq. (\ref{dndt}) can be integrated to yield $N_c(t)$. In particular, for constant $A$ and $V$
\begin{equation}
\label{modello}
\frac{N_c(t)}{N_0}=1 -\exp \left ( -k_d \frac{A}{V} t \right ) ~~~.
\end{equation}
where we estimated $k_d$ to scale with the turbulent kinetic energy of the carrier fluid, $\mathcal{K}_T$, based on the observation that capture is driven by the turbulent fluctuations that transport the particles close to the interface. Ideally, it should be $k_d = C \cdot \mathcal{K}_T^{1/2}$ with $C \simeq 1$:
Through this scaling, the value of $k_d$ can be easily estimated even when RANS-based commercial flow solvers are used.

%

\begin{figure}[t]
 \vspace{0.5cm}
\centering	
	\includegraphics[width=125mm, keepaspectratio, angle=0]{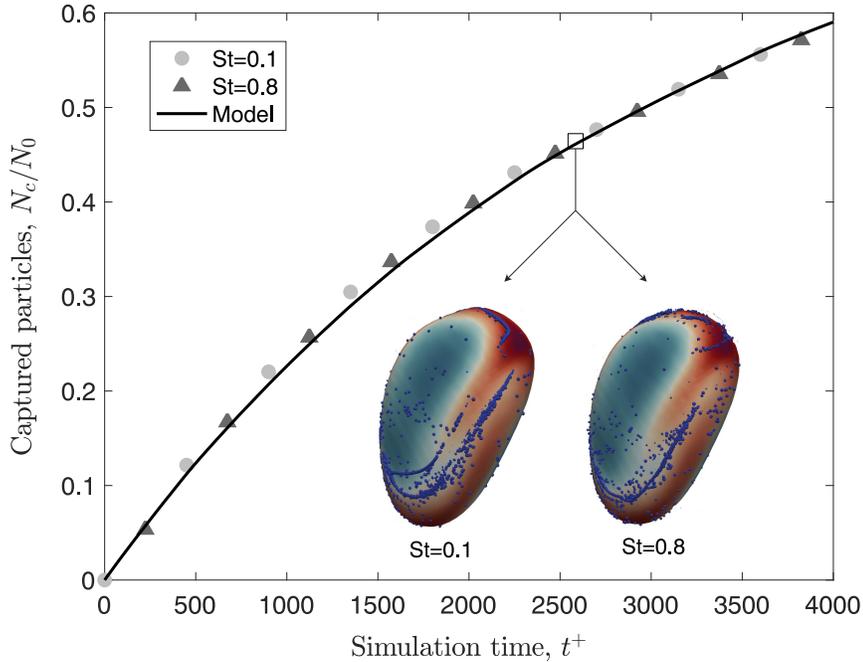}	
			\caption[]{Time evolution of the number of particles captured by the interface. Symbols refer
			to simulation results (circles: $St=0.1$, triangles: $St=0.8$), whereas the solid line refers to
			the model provided by Eq. (\ref{modello}). The inset shows the instantaneous distribution of the 
			captured particles over the interface of a drop, shown in isolation from the flow domain and 	
			colored using the local interface curvature (red: high positive curvature, blue: low negative
			curvature): Particles appear to sample the same interfacial
			regions, confirming the secondary role played by inertia. 
			}
\label{partnum}

%
%

\end{figure}

In Fig. \ref{partnum}, we show the time evolution of $N_c$ obtained from the simulations for the $St=0.1$ and the
$St=0.8$ particles, and we compare numerical results with those yield by Eq. (\ref{modello}). The comparison is proposed
for a dimensionless value of the deposition coefficient that satisfies the $k_d \simeq \mathcal{K}_T^{1/2}$ scaling and for
a dimensionless value of $\mathcal{K}_T$ computed within a fluid layer of thickness equal to 2 wall units around the drop
(rather than over the entire volume occupied by the carrier fluid). This specific thickness corresponds to the volume-averaged
value of the Kolmogorov length scale and represents the last stretch inside the near-interface viscous layer covered by the particles
before capture. In this case, we obtain $\mathcal{K}_T^{1/2} \simeq 0.17$.
%
%
The mean value of $A/V$, also needed in Eq. (\ref{modello}), is equal to $1.3 \cdot 10^{-3}$ at steady state. 
We readily observe that the increase of $N_c$ is unaffected by particle inertia, as one would expect at such
low values of the Stokes number, and follows remarkably well the behaviour predicted by the model.
We remark here that, for the Reynolds number considered in this study, the turbulent kinetic energy averaged over
the entire volume occupied by the fluid is $\langle \mathcal{K}_T \rangle \simeq 1.8$, which yields
$\langle \mathcal{K}_T \rangle^{1/2} \simeq 1.34$ instead of 0.17.
%
This difference can be ascribed to the deformability
of the interface, which acts to damp turbulent fluctuations in the final fluid layer travelled by the particles before
being captured by the interfacial forces. Clearly, using $\langle \mathcal{K}_T \rangle$ instead of $\mathcal{K}_T$ in Eq. (\ref{modello})
would significantly worsen the quantitative agreement with numerical results.
\section{Conclusions}
\label{conclusions}

\vspace{-0.3cm}
In this study, we examine the process of particle capture by large deformable drops in turbulent channel flow, and provide
for the first time a detailed topological characterization of the flow events that control particle adhesion to the drop interface.
To simulate the solid-liquid-liquid three-phase flow, we use a state-of-the-art Eulerian-Lagrangian method based on DNS of turbulence coupled with a Phase Field Model to capture the interface dynamics
and Lagrangian tracking of neutrally-buoyant, sub-Kolmogorov particles. Drops have same density and viscosity
of the carrier liquid, and all three phases are one-way coupled with each other. 
Results discussed in the paper refer to a shear Reynolds number $Re_{\tau} = 150$ and values of the Stokes number
ranging from $St=0.1$ to $St=0.8$. To account for possible modifications due to a change of drop deformability, two values
of the Weber number were considered, $We=0.75$ and $1.5$, but no effect of this parameter was observed.
Therefore, only results relative to $We=0.75$ have been discussed.
An extensive analysis of the topological features of the flow events that drive particle transport toward the surface
of the drops and lead to particle capture has been conducted. By using topology indicators, we were able
to show that particle reach (and adhere to) the interface in regions of positive surface velocity divergence, which are generated by
turbulent fluid motions directed towards the interface. These regions of local flow expansion appear to be well correlated with
high-enstrophy flow topologies that contribute to enstrophy production via vortex compression or stretching.
Fluid motions characterized by high strain and high dissipation are generally avoided by the particles.
An important role is played by the ability of the interface to deform upon interaction with the neighbouring fluid
motions, thus giving rise to highly non-uniform curvature and shear stress distributions. In particular, strong
tangential stresses are produced on the interface, where occurrence of localized elongational flows is favoured.

Based on the topological characterization of the flow seen by the particles during the capture process, a simple
mechanistic model to quantify the fraction of captured particles in time is proposed. This model may be regarded
as a first attempt to lay useful guidelines for the development of physics-aware predictions of transfer rates in
particulate abatement applications, particularly scrubbing.
The proposed model is valid
in the limit of non-interacting particles and exploits the proportionality between the mass flux of particles
that adhere to the interface and the mean concentration of particles that remain afloat in the bulk of the carrier phase: lt is therefore based on a single lumped parameter, the constant of proportionality between flux and concentration.
In spite of its simplicity, the model is capable of reproducing the time increase of the fraction of captured particles with remarkable accuracy when the deposition coefficient is scaled with the turbulent kinetic energy of the fluid measured within one Kolmogorov length scale from the drop. This finding can be explained by the fact that, in the present flow configuration, particle capture is driven by the turbulent fluctuations in the vicinity of the drop interface. For a mechanistic model to work it is therefore necessary to incorporate the effect of these near-interface fluctuations on the overall capture coefficient.

The present work focuses primarily on the process of particle capture.
A future development (which will be the object of an independent study) is therefore the analysis of the dynamics that
characterize the interface-trapped particles as they are driven by both fluid and interfacial stresses. To this aim, it is crucial
to consider a system in which particle-particle collisions are taken into account to reproduce more physically particle
distribution over the interface. Also, the numerical setup should be able to mimic the potential effect of trapped particles on
interface deformability via local modification (reduction) of the surface tension. The surface tension gradients so generated
might produce additional Marangoni stresses on the interface, which might change further the behaviour of trapped particles. 
Other issues to be evaluated are the effects due to density and/or viscosity differences among the phases, which
may induce local modifications of the flow topology in the near-interface regions.

\clearpage

\section*{Acknowledgements}
This work has received funding from the European Union's Horizon 2020 research and innovation programme under the
Marie Sklodowska-Curie grant agreement No. 813948 (COMETE).
The Authors also acknowledge gratefully funding from the PRIN project ``Advanced computations and experiments in turbulent
multiphase flow'' (Project No. 2017RSH3JY).
CINECA (Consorzio inter-universitario per il calcolo automatico, Italy), and VSC (Vienna Scientific Cluster, Austria) are
gratefully acknowledged for the generous allowance of computational resources.

\bibliography{draft} 
\bibliographystyle{unsrt}

\end{document}